%% file: eusipco21.tex
\documentclass[conference]{IEEEtran}
\IEEEoverridecommandlockouts
\usepackage{cite}
\usepackage{amsmath,amssymb,amsfonts, bm}
\usepackage{algorithmic}
\usepackage{graphicx}
\usepackage{textcomp}
\usepackage{xcolor}
\usepackage{pgfplots} 
\usepackage[algoruled,linesnumbered]{algorithm2e}
\usepackage[shortcuts]{glossaries}
\def\BibTeX{{\rm B\kern-.05em{\sc i\kern-.025em b}\kern-.08em T\kern-.1667em\lower.7ex\hbox{E}\kern-.125emX}}
\usetikzlibrary{arrows}
\usetikzlibrary{shapes}
\pgfplotsset{compat=1.14} 

\input{macros.tex}

\SetKwInput{KwInit}{Initialization}
\SetKwInput{KwMain}{Main}
\renewcommand{\baselinestretch}{0.91}

\begin{document}

\title{Low-Complexity Zero-Forcing Precoding for XL-MIMO Transmissions
\thanks{The financial support by the Austrian Federal Ministry for Digital and Economic Affairs, the Austrian National Foundation for Research, Technology and Development, and the Christian Doppler Research Association is gratefully acknowledged.}
}

\author{
    \IEEEauthorblockN{Lucas N. Ribeiro\IEEEauthorrefmark{1}, Stefan Schwarz\IEEEauthorrefmark{3}, Martin Haardt\IEEEauthorrefmark{1}}
    \IEEEauthorblockA{\IEEEauthorrefmark{1}Communications Research Laboratory, Technische Universit\"at Ilmenau, Ilmenau, Germany}
    \IEEEauthorblockA{\IEEEauthorrefmark{3}Christian Doppler Laboratory for Dependable Wireless Connectivity, Technische Universit\"at Wien, Vienna, Austria}
}

\maketitle

\begin{abstract}
    Deploying antenna arrays with an asymptotically large aperture will be central to achieving the theoretical gains of massive MIMO in beyond-5G systems. Such extra-large MIMO (XL-MIMO) systems experience propagation conditions which are not typically observed in conventional massive MIMO systems, such as spatial non-stationarities and near-field propagation. Moreover, standard precoding schemes, such as zero-forcing (ZF), may not apply to XL-MIMO transmissions due to the prohibitive complexity associated with such a large-scale scenario. We propose two novel precoding schemes that aim at reducing the complexity without losing much performance. The proposed schemes leverage a plane-wave approximation and user grouping to obtain a low-complexity approximation of the ZF precoder. Our simulation results show that the proposed schemes offer a possibility for a performance and complexity trade-off compared to the benchmark schemes.
\end{abstract}


\section{Introduction}

Massive \acs{mimo} (\acrlong{mimo}\glsunset{mimo}) is one of the central technologies in the fifth generation of mobile communication systems~\cite{dahlman20205g}. In its current implementation, several antenna elements are compactly arranged at the \ac{bs} to simultaneously serve multiple users. However, an asymptotically large number of antennas is required to effectively achieve the theoretical properties of massive \ac{mimo}. These properties include channel hardening, asymptotic channel orthogonality, among others, and they can be exploited to significantly improve the performance of wireless systems~\cite{marzetta2016fundamentals}.

A natural way to approach the asymptotic massive \ac{mimo} regime consists of deploying an antenna array whose dimensions are orders of magnitude larger than the carrier wavelength. This \ac{mimo} architecture is referred to as \ac{xlmimo} in the literature~\cite{carvalho_non-stationarities_2020}. Potential application scenarios of \ac{xlmimo} include deploying the antenna array on the facade of buildings to or along the city infra-structure~\cite{carvalho_non-stationarities_2020}. A novel aspect of \ac{xlmimo} compared to conventional \ac{mimo} is the presence of visibility regions. Due to the large aperture, portions of the array may not be accessible to some users and the propagation conditions in different visibility regions may be completely uncorrelated, a phenomenon known as spatial non-stationarity. Another crucial difference is that users may not be far apart from the \ac{bs}, such that the exchanged signals experience near-field propagation with spherical wavefronts. To account for these new propagation challenges in \ac{xlmimo}, adapted channel estimation methods~\cite{cheng2019adaptive,han_channel_2020} and novel low-complexity detection schemes~\cite{amiri_extremely_2018, ali_linear_2019, rodrigues_low-complexity_2020} have been proposed. However, little attention has been given to the design of low-complexity precoders for \ac{xlmimo} transmissions yet. 

Classical precoding schemes such as \ac{zf} can be used in \ac{xlmimo} systems to suppress the inter-user interference. However, their requirements in terms of computational resources and \ac{csi} can become prohibitively expensive in such a large-scale scenario. In this work, we propose two novel precoding schemes, namely \ac{mzf} and \ac{tzf}, that aim at solving the complexity problem of \ac{zf}. The proposed \ac{mzf} solution partitions the antenna array into smaller sub-arrays and groups users according to their elevation angles. Based on these sub-arrays, we adopt a plane-wave approximation that allows us to design lower-dimensional precoding filters that approximate the performance of the \ac{zf} precoder, but with much fewer resources. We finally conduct computer simulations to numerically investigate the performance of the proposed schemes.


\emph{Notation:} The transpose, the conjugate transpose, and the pseudo-inverse of a matrix $\bm{X}$ are represented by $\bm{X}^\tran$ and $\bm{X}^\hermit$, $\bm{X}^+$, respectively, and the size of a set $\mathcal{X}$ is $\left|\mathcal{X}\right|$. The $N$-dimensional identity matrix is represented by $\bm{I}_{N}$ and the $(M\times N)$-dimensional null matrix by $\bm{0}_{M\times N}$. The symbol $\delta(\cdot)$ denotes the Kronecker's delta function and $\otimes$ the Kronecker product. The notation $[\bm{v}]_{\mathcal{I}}$ represents the vector obtained by selecting the entries of $\bm{v}$ that corresponds to the index set $\mathcal{I}$.

\section{System Model}

We consider a narrow-band \ac{xlmimo} system in which a transmitter serves $U$ single-antenna users. The transmitter device is equipped with a \ac{ura} of $M = \Mh \cdot \Mv$ antenna elements. Considering the downlink operation, the transmitter applies the precoding filter  $\bm{f}_u \in \mathbb{C}^{M}$ to transmit the data symbol $s_u \in \mathbb{C}$ to each user $u=1,\ldots,U$. Let $\bm{h}_u \in \mathbb{C}^{M}$ denote the downlink channel vector. Then, the received signal by user $u$ can be expressed as 
\begin{equation}
	y_u = \bm{h}_u^\hermit\bm{f}_u s_u + \sum_{j\neq u}^U \bm{h}_u^\hermit \bm{f}_{j} s_j + n_u,
\end{equation}
where $n_u$ denotes a zero mean complex-valued \ac{awgn} component. We assume that $\mathbb{E}[s_us_j^*] = \delta(u-j)$ and $\mathbb{E}[n_u n_j^*] = \sigma_n^2 \cdot \delta(u-j)$. The average \ac{bs} transmit power constraint can be expressed as $\sum_{u=1}^U P_{\text{Tx},u} \leq \Ptx$, with $P_{\text{Tx},u} = \| \bm{f}_u \|_2^2$ denoting the power allocated to user $u$, and $\Ptx \geq 0$ the total transmit power.

\subsection{Channel Model}

\begin{figure}[t]
    \centering
    \includegraphics[scale=1.1]{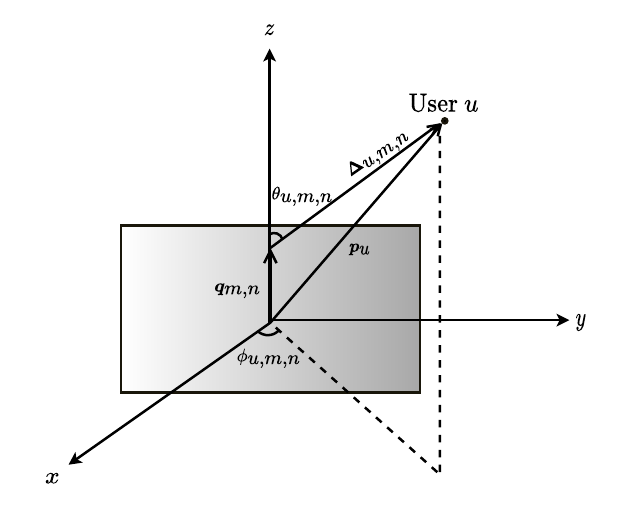}
    \caption{The distance between the $(m,n)$th array element and user $u$ is denoted by $\bm{\Delta}_{u,m,n}=\bm{p}_{u}- \bm{q}_{m,n}$ with azimuth $\phi_{u,m,n}$ and elevation $\theta_{u,m,n}$.}
\end{figure}

We consider that the transmitter antenna array's centroid is placed at the origin, while the single-antenna users are randomly placed around the transmitter. The Cartesian coordinates vector that locates each user is denoted by $\bm{p}_u \in \mathbb{R}^3$, $u=1,\ldots,U$, and the coordinates vector of each array antenna element is represented by 
\begin{equation}
    \bm{q}_{m,n} = \left[0, m-1-\frac{\Mh-1}{2}, n-1-\frac{\Mv-1}{2}\right]^\tran \cdot \frac{\lambda}{2},
\end{equation}
where $\lambda$ denotes the carrier wavelength, $m \in \{1,\ldots,\Mh\}$, and $n \in \{1,\ldots,\Mv\}$. The distance between user $u$ and the $(m,n)$th antenna element in Cartesian coordinates is represented by  $\bm{\Delta}_{u,m,n} = \bm{p}_u - \bm{q}_{m,n}$. This vector can be represented in spherical coordinates as
\begin{subequations}
\begin{align}
    \tilde{\bm{\Delta}}_{u,m,n} &= \left[ r_{u,m,n}, \phi_{u,m,n}, \theta_{u,m,n} \right]^\tran,\\
    r_{u,m,n} &= \| \bm{\Delta}_{u,m,n} \|_2,\\
    \phi_{u,m,n} &= \arctan\left([\bm{\Delta}_{u,m,n}]_2 /[\bm{\Delta}_{u,m,n}]_1\right),\\
    \theta_{u,m,n} &= \arcsin \left([\bm{\Delta}_{u,m,n}]_2/\| \bm{\Delta}_{u,m,n} \|_2\right),
\end{align}
\end{subequations}
where $\phi_{u,m,n}$ and  $\theta_{u,m,n}$ denote the azimuth and elevation angles, respectively. Assuming \ac{los} propagation without multi-paths, the entries of the channel vector $\bm{h}_u$ can be expressed as
\begin{subequations}
\begin{gather} 
    \left[\bm{h}_u\right]_{m + (n-1)\Mh} = \sqrt{\gamma_{u,m,n}} \cdot e^{-j\frac{2\pi}{\lambda}\|\bm{\Delta}_{u,m,n}\|_2}, \label{eq:chan}\\
    m \in \{1,\ldots,\Mh\},\quad n=\{1,\ldots,\Mv\},
\end{gather}
\end{subequations}
with $\gamma_{u,m,n}$ representing the pathloss and antenna gain. 

For later convenience, we define the horizontal and vertical sub-array channel vectors. Let the $p$th horizontal and the $q$th vertical sub-array index sets be respectively defined as 
\begin{subequations}
    \begin{gather}
        \mathcal{I}_{\text{H}}^{(p)} = \{ m + (p-1)\Mh\, | \, m = 1,\ldots, \Mh \},\\
    	\mathcal{I}_{\text{V}}^{(q)} = \{ q + (n-1)\Mh\, | \, n = 1, \ldots, \Mv \},
    \end{gather}
\end{subequations}
for $p \in \{1,\ldots,\Mv\}$ and $q \in \{1,\ldots,\Mh\}$. The respective $p$th horizontal and $q$th vertical subarray channel vectors are given by 
\begin{equation} \label{eq:sub}
    \bm{h}_{\text{H},u,p} = [\bm{h}_u]_{\mathcal{I}_{\text{H}}^{(p)}} \in \mathbb{C}^{\Mh},\quad \bm{h}_{\text{V},u,q} = [\bm{h}_u]_{\mathcal{I}_{\text{V}}^{(q)}} \in \mathbb{C}^{\Mv}.
\end{equation}


In general, for \ac{xlmimo} systems, the angles $\phi_{u,m,n}$ and $\theta_{u,m,n}$ vary over the antenna array, i.e., with different $m,n$. However, if the distances $r_{u,m,n}$ are much larger than the array dimensions $\Mh\cdot \frac{\lambda}{2}$, $\Mv\cdot \frac{\lambda}{2}$, we can adopt the well-established plane-wave approximation~\cite{van2004optimum} to obtain
\begin{gather}
  \phi_{u,m,n} \approx \phi_{u} = \arctan\left([\bm{p}_{u}]_2 /[\bm{p}_{u}]_1\right),\\
  \theta_{u,m,n} \approx \theta_{u} = \arcsin \left([\bm{p}_{u}]_2/\| \bm{p}_{u} \|_2\right),
\end{gather}
and $\gamma_{u,m,n} \approx \gamma_u, \forall m,n$. In this case, the channel vector $\bm{h}_u$ of the \ac{ura} can be approximately written as a Kronecker product of two \ac{ula} vectors corresponding to the horizontal and vertical sub-arrays
\begin{subequations}\label{eq:plane}
\begin{align}
    \bm{h}_u &\approx \sqrt{\gamma_u} \cdot \left(\bm{h}_{\text{V},u} \otimes \bm{h}_{\text{H},u} \right),\\
    \left[\bm{h}_{\text{H},u}\right]_m &= e^{-j\pi \left(m-1-\frac{\Mh-1}{2}\right) \cos\theta_{u} \sin\phi_{u}},\\
    \left[\bm{h}_{\text{V},u}\right]_n &= e^{-j\pi \left(n-1-\frac{\Mv-1}{2}\right) \sin\theta_{u}}, \label{eq:vertical}
\end{align}
\end{subequations}
for $m \in \{1,\ldots,\Mh\}$, and $n \in \{1,\ldots,\Mv\}$. Furthermore, the horizontal and vertical sub-array channel vectors~\eqref{eq:sub} can be approximated as $\bm{h}_{\text{H},u} \approx \bm{h}_{\text{H},u,p}$, and $ \bm{h}_{\text{V},u}\approx \bm{h}_{\text{V},u,q}$, $\forall p,q$.


\section{Precoding Methods}

\subsection{Classical Zero-Forcing (ZF)}

The classical \ac{zf} precoder $\bm{f}_{\text{ZF}, u}$ is designed to satisfy the zero inter-user interference condition
\begin{equation} \label{eq:zf}
	\tilde{\bm{H}}_u \bm{f}_{\text{ZF}, u} = \bm{0}_{(U-1)\times 1},
\end{equation}
where $\tilde{\bm{H}}_u = \left[ \bm{h}_1, \ldots, \bm{h}_{u-1}, \bm{h}_{u+1}, \ldots, \bm{h}_U \right]^\hermit$ denotes the  $(U-1)\times M$-dimensional inter-user interference channel matrix relative to UE $u$. This condition can be satisfied by projecting $\bm{h}_u$ onto the null-space of  $\tilde{\bm{H}}_u$ if $U \leq M$~\cite{marzetta2016fundamentals}. The \ac{zf} precoder is then given by
\begin{gather}\label{eq:zf_bf}
	\bm{f}_{\text{ZF},u} = \frac{\sqrt{P_{\text{Tx},u}}}{\| \tilde{\bm{f}}_u \|_2} \tilde{\bm{f}}_u,\quad \tilde{\bm{f}}_u = (\bm{I}_M - \tilde{\bm{H}}_u^+ \tilde{\bm{H}}_u) \bm{h}_u.
\end{gather}
 
\subsection{Mean-Angle Based Zero-Forcing (MZF)} \label{sec:mzf}
For \ac{xlmimo} systems, the plane-wave approximation~(\ref{eq:plane}) is generally not satisfied, as the antenna array dimensions are in a similar order of magnitude as the user-to-transmitter distances. Hence, to cancel the inter-user interference, the full classical \ac{zf} solution~(\ref{eq:zf_bf}) has to be calculated. For larger arrays and many users, however, this can become prohibitively complex. We therefore propose a low-complexity approximation below, which is based on a plane-wave approximation by partitioning the \ac{ura} into smaller sub-arrays and grouping users accordingly. We consider a partitioning of the \ac{ura} into vertical sub-arrays and a corresponding user grouping in the elevation domain. However, the same approach can also be applied in the horizontal/azimuth domain.

\paragraph{Inter- and intra-group zero-forcing}
The basic idea of the proposed approach is to group users with similar elevation angles, such that interference-cancellation between different groups (inter-group interference) can (approximately) be performed in the elevation domain, whereas interference-cancellation between users of the same group (intra-group-interference) can be performed in the azimuth domain. 

Let $N_g$ be the number of user groups. Group $i \in \{1,\ldots,N_g\}$ contains users $\mathcal{G}_i \subseteq \{1,\ldots,U \},\ \mathcal{G}_i \cap \mathcal{G}_j = \emptyset,\ G_i = \left|\mathcal{G}_i\right|$ and is served from the vertical sub-array consisting of $\Mvi$ consecutive rows of the \ac{ura} indexed by set $\mathcal{I}_{\text{V},i} \subseteq \{1,\ldots,\Mv \},\ \left|\mathcal{I}_{\text{V},i} \right| =\Mvi,\ \mathcal{I}_{\text{V},i} \cap \mathcal{I}_{\text{V},j} = \emptyset$. Group $i$ is thus served from a sub-array consisting of $\Mvi\cdot \Mh$ antenna elements. Notice, this implies a beamforming gain loss of $\Mvi/\Mv$ compared to classical \ac{zf} and \ac{tzf} to be presented in Section~\ref{sec:tzf}.

To perform intra-group-interference cancellation, we assume that the horizontal sub-array channel vector is approximately constant over the $\Mvi$ rows of the sub-array, which is satisfied if the elevation-angle does not vary too much over the sub-array. Specifically, we set $\bar{\bm{h}}_{\text{H},u} = \frac{1}{\Mvi} \sum_{\ell \in \mathcal{I}_{\text{V},i}} \bm{h}_{\text{H},u,\ell}$ and assume $\bm{h}_{\text{H},u,\ell} \approx \bar{\bm{h}}_{\text{H},u}, \forall \ell \in \mathcal{I}_{\text{V},i}$. We then calculate the azimuth beamformer $\bm{f}_{\text{H},u} \in \mathbb{C}^{\Mh}$ of user $u \in \mathcal{G}_i$ to satisfy
\begin{gather}
     \left(\bar{\bm{h}}_{\text{H},j}\right)^\hermit \bm{f}_{\text{H},u} = 0,\ \forall j \in \mathcal{G}_i \setminus u.
\end{gather}
This can be solved similar to~(\ref{eq:zf_bf}), with feasibility condition $G_i \leq \Mh$. Obviously, if the approximation $\bm{h}_{\text{H},u,\ell} \approx \bar{\bm{h}}_{\text{H},u}$ is not satisfied, residual intra-group-interference will occur. 

For inter-group-interference cancellation, we adopt a plane-wave approximation for the elevation angles over the sub-arrays. Specifically, consider sub-array $\mathcal{I}_{\text{V},i}$; we assume $\theta_{u,m,n} \approx \bar{\theta}_{u,i} = \frac{1}{\Mvi\cdot \Mh} \sum_{n \in \mathcal{I}_{\text{V},i}} \sum_{m \in \{1,\ldots,\Mh\}} \theta_{u,m,n}$; notice, in general $\bar{\theta}_{u,i}$ is different for distinct sub-arrays $\mathcal{I}_{\text{V},i}$ and $\mathcal{I}_{\text{V},j}$. Moreover, we assume that the angles of the users within a group are approximately equal $\bar{\theta}_{\ell,i} \approx \bar{\theta}_{i}^{(j)} = \frac{1}{G_j} \sum_{k \in \mathcal{G}_j} \bar{\theta}_{k,i}, \forall \ell \in \mathcal{G}_j$.\footnote{Notice, the two indices $i$, $j$ in $\bar{\theta}_{i}^{(j)}$ are required, since we first average over the sub-array $\mathcal{I}_{\text{V},i}$ and then over users $\mathcal{G}_j$.}  We then calculate the elevation beamformer $\bm{f}_{\text{V},i} \in \mathbb{C}^{\Mvi}$ of group $\mathcal{G}_i$ to satisfy
\begin{gather}
    \left(\bm{h}_{\text{V},i}^{(j)}\right)^\hermit \bm{f}_{\text{V},i} = 0,\ \forall j \in \{1,\ldots,N_g\} \setminus i,
\end{gather}
where $\bm{h}_{\text{V},i}^{(j)} \in \mathbb{C}^{\Mvi}$ is obtained as in~(\ref{eq:vertical}) with elevation angle $\bar{\theta}_{i}^{(j)}$. Again, this can be solved similar to~(\ref{eq:zf_bf}), with feasibility condition $N_g \leq \Mvi$.  Finally, the sub-array beamformer of user $u \in \mathcal{G}_i$ is obtained by the Kronecker product
\begin{gather}
    \bm{f}_{\text{MZF},u} =  \bm{f}_{\text{V},i} \otimes \bm{f}_{\text{H},u},\ \bm{f}_{\text{MZF},u} \in \mathbb{C}^{\Mvi\cdot \Mh}.
\end{gather}

\paragraph{User grouping and array partitioning}
The separable beamforming approach described above imposes several assumptions and conditions that should be satisfied to avoid excessive residual inter-user interference. This can be assured by appropriate user grouping and array partitioning. 

First of all, we have to satisfy the feasibility conditions $G_i \leq \Mh,\ N_g \leq \Mvi,\  \forall i \in \{1,\ldots,N_g\}$ to enable inter- and intra-group-interference cancellation. In addition, the variation of elevation angles amongst users of the same group, as well as, the variation of each user's elevation angles over the antenna elements of a sub-array should be sufficiently small to validate the mean-angle based channel vector approximations. We propose a greedy approach to achieve these targets: 
\begin{enumerate}
    \item user grouping under the assumption $\Mvi = \lfloor\frac{\Mv}{N_g}\rfloor, \forall i$;
    \item optimization of the array partitioning given fixed user groups.
\end{enumerate}

The proposed greedy user grouping approach is summarized in Algorithm\,\ref{alg:grouping}. In this algorithm, we group users with small angular distances using a distance threshold $\theta_t$. We increase $\theta_t$ within the algorithm until the number of obtained groups $N_g$ is sufficiently small to satisfy the feasibility condition $N_g < \frac{\Mv}{N_g}$, assuming equal sub-array partitioning.   

\begin{algorithm}[t]
	\small{
 \KwIn{Average user angles $\theta_u$, grouping threshold $\theta_t$
}
\KwMain{\\}
\Repeat{$N_g < \frac{\Mv}{N_g}$}{
Initialize group counter $i = 0$\\
Initialize set of ungrouped users $\bar{\mathcal{G}} = \{1,\ldots,U\}$ \\
\Repeat{$\bar{\mathcal{G}} = \emptyset$}{
Increase group counter $i = i+1$\\
Sort set $\bar{\mathcal{G}}$ according to increasing angle $\theta_u$ \\
Find user $u \in \bar{\mathcal{G}}$ with smallest angular distance to neighbouring users\\
Group user $u$ with at most $(\Mh-1)$ closest neighbors with angular distance less than $\theta_t$ in $\mathcal{G}_i$\\ 
Update set of ungrouped users $\bar{\mathcal{G}} = \bar{\mathcal{G}} \setminus \mathcal{G}_i$
}
Set number of groups $N_g = i$\\
Increase grouping threshold $\theta_t = 2\cdot \theta_t$
}
\KwOut{
	User groups $\mathcal{G}_i, \forall i \in \{1,\ldots,N_g\}$
}
 \caption{\small{Elevation angle based user grouping.}}
 \label{alg:grouping}
}
\end{algorithm}

Depending on the elevation angles of the users, the proposed greedy user grouping may potentially lead to strongly unbalanced group sizes. To compensate for this, we optimize in a second step the sub-array sizes $\Mvi$, attempting to achieve similar ratios $\Mvi/G_i, \forall i$.

\vspace{-0.3cm}
{\small
\[
    \max_{\Mvi \in \mathbb{N}, \forall i}\ \  \min_{i \in \{1,\ldots,N_g\}} \frac{\Mvi}{G_i}\quad
    \text{s.t. }\,
    \Mvi \geq N_g,\ \sum_{i = 1}^{N_g} \Mvi \leq \Mv.
\]}
This linear integer programming problem can be solved to optimality by an appropriate integer programming solver, or it can be solved approximately by standard integer-relaxation techniques~\cite{Wolsey1999}.

\subsection{Tensor Zero-Forcing (TZF)} \label{sec:tzf}

The \ac{tzf} precoder aims at satisfying the zero inter-user interference~\eqref{eq:zf} by exploiting some algebraic properties of bi-dimensional \ac{los} channels~\cite{ribeiro2020lowcomplexity}. This precoder is able to approximate the interference cancellation performance of classical \ac{zf} with much less stringent \ac{csi} and computational requirements. 

The \ac{tzf} precoder adopts the plane-wave approximation~\eqref{eq:plane} and assumes that all scheduled users have approximately the same elevation (or azimuth) angles. Let 
\begin{gather}
	\tilde{\bm{H}}_{\text{H},u} = \left[ \bm{h}_{\text{H},1}, \ldots, \bm{h}_{\text{H},{u-1}}, \bm{h}_{\text{H},{u+1}}, \ldots, \bm{h}_{\text{H},U} \right]^\hermit,\\
	\tilde{\bm{H}}_{\text{V},u} = \left[ \bm{h}_{\text{V},1}, \ldots, \bm{h}_{\text{V},{u-1}}, \bm{h}_{\text{V},{u+1}}, \ldots, \bm{h}_{\text{V},U} \right]^\hermit. \label{eq:mui_v}
\end{gather}
denote the horizontal and vertical inter-user interference channel matrices, respectively. The \ac{tzf} precoder is given by~\cite{ribeiro2020lowcomplexity}
\begin{subequations}
    \begin{gather}
        \bm{f}_{\text{TZF},u} = \frac{\sqrt{P_{\text{Tx},u}}}{\| \bar{\bm{f}}_u \|_2} \bar{\bm{f}}_u,\\
        \bar{\bm{f}}_u = \left[ \bm{I}_{M} - (\Pv \otimes \Ph) \right] (\bm{h}_{\text{V},u} \otimes \bm{h}_{\text{H},u}), \label{eq:tzfpart}
    \end{gather}
\end{subequations}
with $\Pv = \tilde{\bm{H}}_{\text{V},u}^+\tilde{\bm{H}}_{\text{V},u}$ and $\Ph = \tilde{\bm{H}}_{\text{H},u}^+\tilde{\bm{H}}_{\text{H},u}$ representing projectors onto the row-space of $\tilde{\bm{H}}_{\text{V},u}$ and $\tilde{\bm{H}}_{\text{H},u}$, respectively.

If the equal elevation angles assumption is satisfied, then the rows of the vertical inter-user interference matrix \eqref{eq:mui_v} become highly collinear, and the vertical-domain processing does not improve the interference cancellation. We leverage this fact to further simplify the \ac{tzf} precoder by applying only horizontal-domain processing. In this case, we set $\Pv = \bm{I}_{\Mv}$ and further express \eqref{eq:tzfpart} as
\begin{subequations} \label{eq:tzf}
\begin{align}
    \bar{\bm{f}}_u &= (\bm{h}_{\text{V},u} \otimes \bm{h}_{\text{H},u}) - [\bm{h}_{\text{V},u} \otimes (\Ph\bm{h}_{\text{H},u})] \\
    &= \bm{h}_{\text{V},u} \otimes \left[ (\bm{I}_{\Mh} - \Ph) \bm{h}_{\text{H},u}\right].
\end{align}
\end{subequations}
Equation~\eqref{eq:tzf} reveals that the \ac{tzf} precoder can be seen as the Kronecker (tensor) product between a horizontal-domain \ac{zf} precoder and a vertical-domain \ac{mrt} precoder. Since the interference cancellation is performed solely in the horizontal domain, the feasibility condition is then $U \leq \Mh$.

\subsection{Complexity Analysis}

The classical \ac{zf} precoder requires instantaneous \ac{csi} from all $U$ users and \eqref{eq:zf_bf} can be calculated by $O(\Mh^3 \cdot \Mv^3)$ operations. By contrast, \ac{mzf} requires only partial \ac{csi} (horizontal sub-array channels $\bm{h}_{\text{H},u,\ell}$ and the elevation angles $\theta_{u,m,n}$) and its beamforming filters can be computed by $O(\Mh^3) + O(\Mvi^3)$. Likewise, \ac{tzf} demands only partial \ac{csi} (horizontal sub-array channels $\bm{h}_{\text{H},u,\ell}$) and its precoding filters can be obtained by $O(\Mh^3)$ operations. It is clear from this short analysis that, in \ac{xlmimo} systems, \ac{mzf} and \ac{tzf} are significantly less complex than the standard \ac{zf} solution.

\section{Simulation Results}

In this section, we present the computer simulation experiments designed to analyze the performance of proposed precoding schemes. The proposed precoding schemes exploit the plane-wave approximation~\eqref{eq:plane} and assume that the elevation angles of different users are approximately the same or they are spread around clusters. To investigate the precoders' sensibility to these assumptions, we randomly place the users around the transmitter in a way that allow us to control the assumptions' plausibility. Specifically, we randomly generate the spherical coordinates $\bm{p}_u = [r_u, \phi_u, \theta_u]^\tran$ of each user $u$ as follows. 
\begin{itemize}
    \item The radial coordinate $r_u$ is sampled from a uniform random variable distributed in $[d, 2d]$, where $d$ is a  parameter that controls how far the user is placed from the transmitter's antenna array;
    \item The azimuth angle $\phi_u$ is sampled from a uniform random variable defined in $[-s_{\text{az}}, s_{\text{az}}]$, with $s_{\text{az}}$ representing the azimuth spread angle;
    \item To generate the elevation angles, the $U$ users are first divided into $N_c$ groups of $\lceil U/N_c \rceil$ users. Group $g \in \{1,\ldots,N_c\}$ contains users $\mathcal{L}_g \subset \{ 1,\ldots, U \}$, $\mathcal{L}_i \cap \mathcal{L}_j = \emptyset$ and is associated with a mean elevation cluster angle $\mu_g$ and intra-cluster elevation spread $\sigma_{g}$. The elevation angle $\theta_u$ of user $u \in \mathcal{L}_g$ is therefore sampled from a Gaussian distribution with mean $\mu_g$ and standard deviation $\sigma_g$.
\end{itemize} 

It is important to recall that the \ac{mzf} and \ac{tzf} precoders have different assumptions concerning the elevation angles of the scheduled users. As described in Section~\ref{sec:mzf}, \ac{mzf} allows inter-group angular variation but it expects small intra-group angular variation. By contrast, \ac{tzf} assumes that the elevation angles of all scheduled users are approximately the same. Therefore, to fairly compare these precoding schemes, \ac{mzf} schedules all $U$ users in the same time-frequency resources elements, whereas \ac{tzf} employs orthogonal scheduling to serve user groups $\mathcal{G}_i$ with the same elevation angles on different resource elements.

The figures of merit considered in our simulations are the \ac{sinr} and the achievable sum-rate. We calculate the \ac{sinr} of user $u$ in group $i$ as
\[
    \text{SINR}_{u,i} = \frac{|\bm{h}_u^\hermit \bm{f}_u|^2}{ \sum_{\substack{p \in \mathcal{G}_i \\ p\neq u}} |\bm{h}_u^\hermit \bm{f}_p|^2  + \sum_{\ell\neq i}^{N_g} \sum_{q \in \mathcal{G}_\ell} |\bm{h}_u^\hermit \bm{f}_q|^2 + \sigma_n^2 }.
\]
The first term in the denominator represents the intra-group interference, while the second term the inter-group interference. Notice that the inter-group interference term is zero when orthogonal scheduling is employed. The achievable sum-rate can be calculated as 
\begin{equation} \label{eq:sr}
    \text{SR} = \sum_{i=1}^{N_g} \sum_{u \in \mathcal{G}_i} \log_2 \left( 1 + \text{SINR}_{u,i}\right).
\end{equation}
In case of orthogonal scheduling, the achievable sum-rate \eqref{eq:sr} is normalized as $\overline{\text{SR}} = \text{SR}/N_g$ to account for the time-sharing loss.

The following parameters were considered in the performed simulation experiments. The transmitter antenna array contains $M=2000$ elements with $(\Mh, \Mv)=(50, 40)$ to serve $U=20$ single-antenna users in total. The effects of the individual antenna gain and path-loss are not regarded in the reported simulations, i.e., $g_{u,m,n}=1$, $\forall u,m,n$. The carrier frequency is $2$~GHz, the \ac{awgn} noise power is set to $\sigma_n^2=10^{-2}$, and the initial grouping threshold is set to $2^\circ$. Furthermore, the mean elevation cluster angles $\mu_g$, $g\in \{1,\ldots,N_c\}$, are sampled from a uniform random variable defined in  $[-s_{\text{el}},s_{\text{el}}]$, where $s_{\text{el}}$ represents the elevation inter-cluster angle spread. Unless stated otherwise, the azimuth spread is set to $s_{\text{az}} = 60^\circ$, the intra-cluster elevation spread to $\sigma_g=1^\circ$, $\forall g$, and the inter-cluster elevation spread to $s_{\text{el}} = 60^\circ$. The results reported in the section were obtained from $1000$ independent experiments.

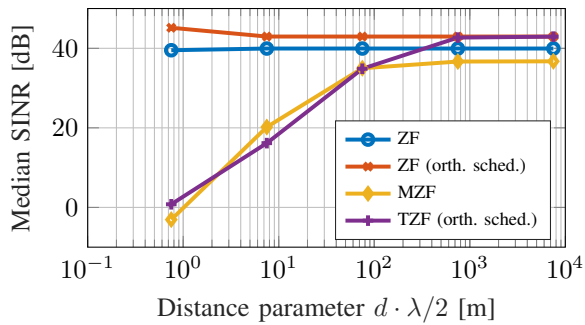
\begin{figure}[t]
	\centering
	\input{./exp1.tex}
	\caption{Median \ac{sinr} performance as a function of the distance parameter $d$, $s_{\text{az}}=s_{\text{el}}=60^\circ$, $\sigma_g=1^\circ\, \forall g$, $N_c=2$ elevation clusters.}
	\label{fig:dist}
\end{figure}

In the first experiment, we evaluate the effect of the distance between users and transmitter on the \ac{sinr} performance. Figure~\ref{fig:dist} depicts the median \ac{sinr} as a function of the parameter $d$ that determines the users' radial coordinate. We observe that \ac{mzf} and \ac{tzf} are quite sensitive to this parameter, while the standard \ac{zf} method exhibits some robustness. As we do not consider pathloss, the \ac{sinr} degradation observed in the proposed methods is explained by the plane-wave approximation. The closer the users are to the transmitter, the less accurate the plane-wave approximation \eqref{eq:plane} is. As a consequence, \ac{mzf} and \ac{tzf} are not able to properly cancel the inter-user interference. We also observe that the solutions employing orthogonal scheduling provide larger \ac{sinr} values. In this case, fewer users are scheduled together in the same time-frequency resources, hence less interference is produced in the considered time-frequency resource elements.

\begin{figure}[t]
	\centering
	\input{./exp2.tex}
	\caption{Median \ac{sinr} performance as a function of the intra-cluster elevation spread $\sigma$,  $s_{\text{az}}=s_{\text{el}}=60^\circ$, $d\cdot \tfrac{\lambda}{2}=750$ meters, $N_c=2$ elevation clusters.}
	\label{fig:angles}
\end{figure}
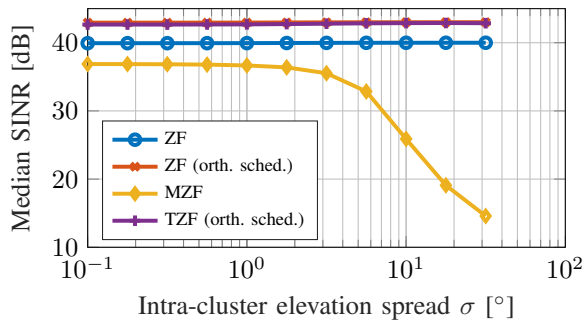

\begin{figure}[t]
	\centering
	\input{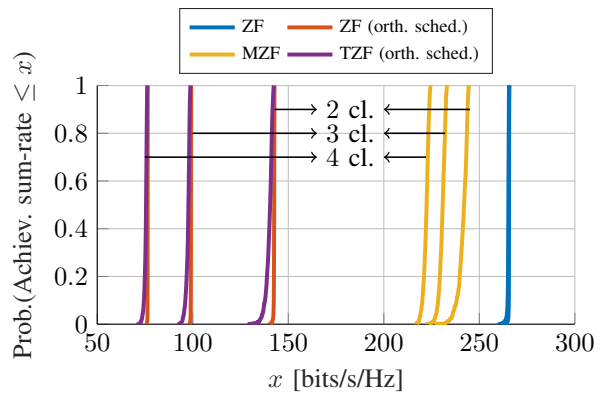}
	\caption{Achievable sum-rate empirical \acs{cdf}, $s_{\text{az}}=s_{\text{el}}=60^\circ$, $\sigma_g=1^\circ\, \forall g$, $d\cdot \tfrac{\lambda}{2}=750$ meters, $N_c\in \{2,3,4\}$ elevation clusters.}
	\label{fig:sumrate}
\end{figure}

In the second experiment, we assess the influence of the intra-cluster elevation spread $\sigma_g$ on the \ac{sinr} performance. The median \ac{sinr} is plotted as a function of the intra-cluster elevation spread in Figure~\ref{fig:angles}. In this experiment, the intra-cluster elevation spreads $\sigma_g$, $g\in \{1,\ldots,N_c\}$, are set to the same value $\sigma$, i.e., $\sigma_1 = \cdots = \sigma_{N_c}= \sigma$. Moreover, the product $d \cdot \tfrac{\lambda}{2}$ is set to $750$ meters. As we can see in Figure~\ref{fig:dist}, the plane-wave approximation holds for such a distance. Figure~\ref{fig:angles} indicates that \ac{mzf} is sensitive to the intra-cluster elevation spread parameter. This is because the inter-group-interference cancellation of \ac{mzf}, which relies on the mean elevation angles, becomes less accurate with the increase of $\sigma$, causing residual inter-user interference. By contrast, \ac{tzf} exhibits more robustness to the increase of the intra-cluster elevation spread. Although large spreading values violate the equal elevation angles assumption, these results indicates that the azimuth-domain based interference cancellation is enough to mitigate the inter-user interference.

In the final experiment, we investigate the achievable sum-rate performance of the proposed methods for different number of user clusters. The target of this experiment is to provide insights into the trade-off between the beamforming gain losses of \ac{mzf} and the orthogonal scheduling time-sharing losses. In Figure~\ref{fig:sumrate}, we plot the achievable sum-rate empirical \ac{cdf} for $d\cdot \tfrac{\lambda}{2} = 750$ meters, intra-cluster elevation spread of  $\sigma = 1^\circ$, and number of clusters $N_c \in \{2,3,4\}$. This figure indicates that the throughput of the orthogonal scheduling-based precoders and \ac{mzf} tend to decrease with the increase of the number of clusters. When this number increases, the orthogonal scheduling based precoders need to use more time-frequency resources, therefore reducing the spectral efficiency. Furthermore, \ac{mzf} tends to form smaller sub-arrays to keep the angular variation low when the number of user cluster increases. Since the sub-array dimensions decrease, the beamforming gain is smaller, reducing the throughput.

\section{Conclusion}

In this work, we propose novel precoding schemes for \ac{xlmimo} transmissions that aim at solving the complexity issue of classical precoding schemes, such as \ac{zf}. The proposed \ac{mzf} and \ac{tzf} solutions resort to a plane-wave approximation to partition the transmitter's array into smaller sub-arrays and to group users according to their elevation angles, thereby allowing to approximately factorize the \ac{zf} filter into a Kronecker product. Such a factorization significantly reduces the \ac{csi} and computational requirements as compared to the classical \ac{zf} precoder. Our simulation results show that the proposed schemes are capable of well-approximating the benchmark solutions. The performance gap gets tighter as the plane-wave approximation becomes more accurate and when the intra-cluster elevation spread decreases. We also notice that the beamforming gain is reduced as the number of user clusters increases. By carefully scheduling the users, it is possible to reduce the number of elevation clusters and the intra-cluster elevation spread, offering a possibility for a performance and complexity trade-off.

\bibliographystyle{IEEEtran}
\bibliography{./eusipco21}

\end{document}

%% file: macros.tex
\newcommand{\Mh}{\ensuremath{M_{\text{H}}}}
\newcommand{\Mv}{\ensuremath{M_{\text{V}}}}
\newcommand{\Mvi}{\ensuremath{M_{\text{V},i}}}
\newcommand{\Ptx}{\ensuremath{P_{\text{Tx}}}}
\newcommand{\Ph}{\ensuremath{\bm{P}_{\text{H}}}}
\newcommand{\Pv}{\ensuremath{\bm{P}_{\text{V}}}}

\newacronym{bs}{BS}{base station}
\newacronym{tdd}{TDD}{time division duplexing}
\newacronym{tti}{TTI}{transmission time interval}
\newacronym{cdf}{cdf}{cumulative distribution function}
\newacronym{los}{LOS}{line of sight}
\newacronym{nlos}{NLOS}{non-line of sight}
\newacronym{mimo}{MIMO}{multiple-input multiple-output}
\newacronym{ula}{ULA}{uniform linear array}
\newacronym{ura}{URA}{uniform rectangular array}
\newacronym{zmcsg}{ZMCSG}{zero mean circularly symmetric Gaussian}
\newacronym{awgn}{AWGN}{additive white Gaussian noise}
\newacronym{csi}{CSI}{channel state information}
\newacronym{snr}{SNR}{signal to noise ratio}
\newacronym{sir}{SIR}{signal to interference ratio}
\newacronym{rf}{RF}{radio-frequency}
\newacronym{dft}{DFT}{discrete Fourier transform}
\newacronym{idft}{IDFT}{inverse discrete Fourier transform}
\newacronym{ls}{LS}{least squares}
\newacronym{xlmimo}{XL-MIMO}{extra-large MIMO}
\newacronym{zf}{ZF}{zero-forcing}
\newacronym{tzf}{TZF}{tensor zero-forcing}
\newacronym{mrt}{MRT}{maximum-ratio transmission}
\newacronym{mzf}{MZF}{mean-angle based zero-forcing}
\newacronym{sinr}{SINR}{signal to interference and noise ratio}
\newacronym{irs}{IRS}{intelligent reflective surface}

\newcommand{\tran}{\mathsf{T}}						
\newcommand{\hermit}{\mathsf{H}}					

%% file: exp1.tex
%
%
\definecolor{mycolor1}{rgb}{0.00000,0.44700,0.74100}%
\definecolor{mycolor2}{rgb}{0.85000,0.32500,0.09800}%
\definecolor{mycolor3}{rgb}{0.92900,0.69400,0.12500}%
\definecolor{mycolor4}{rgb}{0.49400,0.18400,0.55600}%
\begin{tikzpicture}

\begin{axis}[%
width=2.5in,
height=1.25in,
scale only axis,
xmode=log,
xmin=0.1,
xmax=10000,
xminorticks=true,
xlabel style={font=\color{white!15!black}},
xlabel={Distance parameter $d \cdot \lambda/2$ [m]},
ymin=-10,
ymax=50,
ylabel style={font=\color{white!15!black}},
ylabel={Median SINR [dB]},
axis background/.style={fill=white},
xmajorgrids,
xminorgrids,
ymajorgrids,
legend style={at={(0.97,0.03)}, anchor=south east, legend cell align=left, align=left, draw=white!15!black, font=\scriptsize}
]
\addplot [color=mycolor1, line width=1.5pt, mark=o, mark options={solid, mycolor1}]
  table[row sep=crcr]{%
0.75	39.516406949213\\
7.5	39.9435240589249\\
75	39.9447765814006\\
750	39.944906262484\\
7500	39.9444341845902\\
};
\addlegendentry{ZF}

\addplot [color=mycolor2, line width=1.5pt, mark=x, mark options={solid, mycolor2}]
  table[row sep=crcr]{%
0.75	45.1666905899581\\
7.5	42.9604045260857\\
75	42.9554984014355\\
750	42.9557567346203\\
7500	42.9552913079364\\
};
\addlegendentry{ZF (orth. sched.)}

\addplot [color=mycolor3, line width=1.5pt, mark=diamond, mark options={solid, mycolor3}]
  table[row sep=crcr]{%
0.75	-3.07628061730736\\
7.5	20.2026941267261\\
75	34.9861764946619\\
750	36.6740733095804\\
7500	36.7268169287668\\
};
\addlegendentry{MZF}

\addplot [color=mycolor4, line width=1.5pt, mark=+, mark options={solid, mycolor4}]
  table[row sep=crcr]{%
0.75	0.781946818468717\\
7.5	16.1579252685316\\
75	34.8609227618848\\
750	42.6923942701272\\
7500	42.9351528703732\\
};
\addlegendentry{TZF (orth. sched.)}

\end{axis}

\end{tikzpicture}%

%% file: exp2.tex
%
%
\definecolor{mycolor1}{rgb}{0.00000,0.44700,0.74100}%
\definecolor{mycolor2}{rgb}{0.85000,0.32500,0.09800}%
\definecolor{mycolor3}{rgb}{0.92900,0.69400,0.12500}%
\definecolor{mycolor4}{rgb}{0.49400,0.18400,0.55600}%
\begin{tikzpicture}

\begin{axis}[%
width=2.5in,
height=1.25in,
scale only axis,
xmode=log,
xmin=0.1,
xmax=100,
xminorticks=true,
xlabel style={font=\color{white!15!black}},
xlabel={Intra-cluster elevation spread $\sigma$ [$^\circ$]},
ymin=10,
ymax=45,
ylabel style={font=\color{white!15!black}},
ylabel={Median SINR [dB]},
axis background/.style={fill=white},
xmajorgrids,
xminorgrids,
ymajorgrids,
legend style={at={(0.03,0.03)}, anchor=south west, legend cell align=left, align=left, draw=white!15!black,font=\scriptsize}
]
\addplot [color=mycolor1, line width=1.5pt, mark=o, mark options={solid, mycolor1}]
  table[row sep=crcr]{%
0.1	39.9327630910382\\
0.177827941003892	39.9334722393734\\
0.316227766016838	39.9341578711496\\
0.562341325190349	39.9369453313839\\
1	39.9444326626416\\
1.77827941003892	39.9599290764365\\
3.16227766016838	39.9773288797902\\
5.62341325190349	39.9878123489015\\
10	39.9938404965247\\
17.7827941003892	39.9966396478791\\
31.6227766016838	39.9974701598722\\
};
\addlegendentry{ZF}

\addplot [color=mycolor2, line width=1.5pt, mark=x, mark options={solid, mycolor2}]
  table[row sep=crcr]{%
0.1	42.9436912957729\\
0.177827941003892	42.9443412691044\\
0.316227766016838	42.944944610917\\
0.562341325190349	42.9479278873669\\
1	42.9552929273019\\
1.77827941003892	42.9708787003467\\
3.16227766016838	42.988185805959\\
5.62341325190349	42.9987312021972\\
10	43.004805801357\\
17.7827941003892	43.0078002543803\\
31.6227766016838	43.0087520196354\\
};
\addlegendentry{ZF (orth. sched.)}

\addplot [color=mycolor3, line width=1.5pt, mark=diamond, mark options={solid, mycolor3}]
  table[row sep=crcr]{%
0.1	36.8759937315694\\
0.177827941003892	36.8650871645358\\
0.316227766016838	36.8408954906381\\
0.562341325190349	36.7882566533866\\
1	36.6668999312779\\
1.77827941003892	36.3844194937012\\
3.16227766016838	35.5246328967341\\
5.62341325190349	32.8535835490848\\
10	25.8852587738086\\
17.7827941003892	19.0841356834725\\
31.6227766016838	14.5820152677189\\
};
\addlegendentry{MZF}

\addplot [color=mycolor4, line width=1.5pt, mark=+, mark options={solid, mycolor4}]
  table[row sep=crcr]{%
0.1	42.6651016657526\\
0.177827941003892	42.665453370268\\
0.316227766016838	42.6608977143295\\
0.562341325190349	42.676128037465\\
1	42.6887380256875\\
1.77827941003892	42.7352166883421\\
3.16227766016838	42.7897311706798\\
5.62341325190349	42.84558785141\\
10	42.874005844729\\
17.7827941003892	42.8898081315603\\
31.6227766016838	42.8635109935329\\
};
\addlegendentry{TZF (orth. sched.)}

\end{axis}

\end{tikzpicture}%